\documentclass[12pt,prl,aps]{revtex4}
\usepackage[dvips]{graphicx}

\begin{document}
\title{Brazilian elections: voting for a scaling democracy}
\author{R. N. Costa Filho${^*}$, M. P. Almeida, J. E. Moreira, J. S.
Andrade Jr.}
\affiliation{Departamento de F\'{\i}sica, Universidade Federal
do Cear\'a,Caixa Postal 6030\\ Campus do Pici, 60451-970 Fortaleza,
Cear\'a,Brazil}

\begin{abstract}
The proportional elections held in Brazil in 1998 and 2002 display identical
statistical signatures. In particular, the distribution of votes among
candidates includes a power-law regimen. We suggest that the rationale behind
this robust scaling invariance is a multiplicative process in which the voter's
choice for a candidate is governed by a product of probabilities.
\end{abstract}
\maketitle

The election is a fundamental democratic process and the vote certainly
represents the most effective instrument for regular citizens to promote
significant changes in their communities. General elections in Brazil are
held every four years, when citizens vote for executive (president and
state governors) as well as legislative (congressmen and state deputies)
mandates. Voting is compulsory and ballots are collected in electronic
voting machines. Previously, we reported a statistical analysis of the
Brazilian 1998 elections showing that the proportional voting process for
state deputies displayed scale invariance \cite{Costa}. It was shown that
the distribution of the number of candidates $N$ receiving a
fraction of votes $v$ followed a power-law $N(v)\sim v^\alpha$,
where $\alpha\approx-1$, extending over two orders of magnitude. The
striking similarity in the distribution of votes in all states, regardless
of large diversities in social and economical conditions in different
regions of the country, was taken as an indication of a common mechanism in
the decision process. The issue was raised, however, that this uniformity
in behavior could be due to some peculiarity of the political rules valid
at that moment. In 1998, the legislation allowed party alliances of all
sorts, not necessarily reproduced from state to state and independent from
the majority candidatures. Since then, the National Congress has approved
new political rules. Any alliance of parties supporting a presidential
candidate should be reproduced in all states with restrictions on the
campaigns for Congress and local representation.

On October 6, 2002, 115,253,432 electors from Brazil's 27 states chose from
among 4,210 candidates for federal and 11,717 for state deputies.  The
collection of ballots was entirely electronic, thus permitting a very rapid
count and publication of the results \cite{tse}. We apply the same
statistical treatment given to the results of the 1998 election for congressmen
and state deputies. Votes for the candidates are normalized by the number of
voters in their respective states. We then rank the candidate by his/her
normalized number of votes and perform the statistics for the whole country.
This is justified by the aforementioned compatibility found in the results of
1998 for each state, and verified again for the 2002 election.

Figures 1(a) and (b) are the log-log plots of the voting distribution
$N(v)$ using data of all candidates in Brazil competing for
positions of state and federal deputies, respectively. The voting
distributions for 1998 and 2002 are almost identical for both types of
candidates. Furthermore, in all cases we can identify a well-defined region
where the voting distributions follow a $1/v$ behaviour. The
remarkable similarity between the statistics of the 1998 and 2002 results,
despite significant modifications of the electoral rules, indicates that
the processes behind a voting decision are largely insensitive to external
factors. We argue that this is a manifestation of a multiplicative relation
of independent factors influencing the choice of the elector. The voting
fraction $v$ of a candidate can be viewed as a ``grand-process''
depending on the successful completion of a number $n$ of
independent ``sub-processes''. Each sub-process could be associated with
the characteristics of a given candidate or with his/her position on a
particular political, social or economical issue. We can then link to each
candidate a probability $p_i$ of performing sub-process $i$, so that
his/her voting fraction is  $v=p_1p_2\cdots p_n$. Assuming that the $p_i$'s
are independent random variables, and considering $n$ sufficiently
large we find, from the Central Limit Theorem, that the distribution of
$v$ should be approximately log-normal. If the statistical
dispersion of the data is large enough, the log-normal distribution can
``mimic'' the $1/v$ profile over a given range of
values \cite{west}. Deviations at small and large values of $v$ are
expected to be due to finite size effects.

The statistical analyses of the 1998 and 2002 Brazilian elections indicate that
the voting distribution is highly reproducible. It displays features of a
``scale-free'' phenomenon \cite{mant,stan,vian} where the governing
decision-making mechanism can be adequately modeled in terms of a
multiplicative process \cite{west}. A candidate for the National
Congress or for the State Houses is likely to be known by the voters
through the media. When instruments of information are absent or ignored,
the voting decision is probably determined by direct communication with
friends, relatives, etc.  In the case of municipal elections, for instance,
the physical proximity of candidates and voters creates a different
framework from the one found for state and federal elections. We performed
the same data analysis to the results of Brazilian municipal elections held
in 2000. No clear indication of power-law behaviour was found even for the
most populated cities.

The authors are supported by the Brazilian National
Research Agency CNPq.

\newpage
\begin{figure}
\includegraphics[width=10cm, height = 9cm]{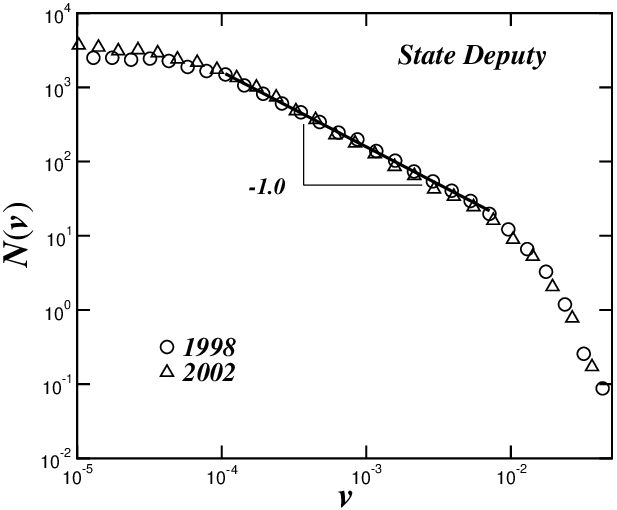}

\caption{}
\end{figure}

\vspace {4.0cm}
\begin{figure}

\includegraphics[width=10cm, height = 9cm]{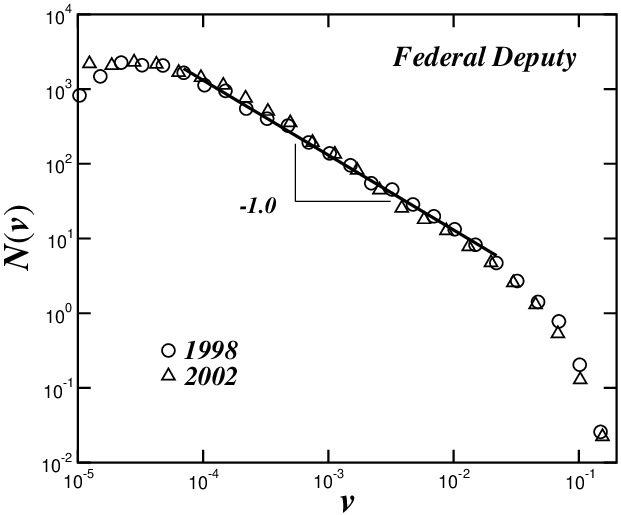}
\caption{(a) Double logarithmic plot of the voting distribution for state deputies in
1998 (circles) and 2002  (triangles). The solid lines are the least-squares fits
to the data in the scaling regions. The numbers indicate the scaling exponent .
(b) The same as in (a) but for federal deputies.}
\end{figure}


\begin{thebibliography}{cc}
\bibitem[*]{rncf} Email address : rai@fisica.ufc.br

\bibitem{Costa} R. N. Costa Filho, M. P. Almeida, J. S. Andrade Jr., J. E.
Moreira, Phys. Rev. E {\bf 60}, 1067 (1999).

\bibitem{tse} http://www.tse.gov.br

\bibitem{west} B. J. West, M. F. Shlesinger, Int. J. Mod. Phys. B {\bf 3},
795 (1989).

\bibitem{mant} R. N. Mantegna, H. E. Stanley, Nature
{\bf 376}, 46 (1995).

\bibitem{stan} M. H. R. Stanley, L. A. N. Amaral, S. Buldyrev, S. Havlin,
H. Leschhorn, P. Maass, M. A. Salinger, and H. E. Stanley, Nature
{\bf 379}, 804 (1996).

\bibitem{vian} P. C. Ivanov, M. G. Rosenblum, C. K. Peng, J. Mietus,
S. Havlin, H. E. Stanley, A. L. Goldberger, Nature
{\bf 383}, 323 (1996).


\end{thebibliography}
\end{document}